\documentclass[twoside,12pt]{article}
\usepackage{epsfig}
\usepackage{bigstrut}
\usepackage{amssymb}
\usepackage{amsmath}

\def\Journal#1#2#3#4{{#1} {#2} (#4) #3 }

\def\NPA{{\em Nucl. Phys.} A}

\def\NPB{{\em Nucl. Phys.} B}

\def\PLB{{\em Phys. Lett.} B}

\def\PRL{\em Phys. Rev. Lett.}

\def\PRD{{\em Phys. Rev.} D}
\def\PRC{{\em Phys. Rev.} C}

\def\ZPC{{\em Z. Phys.} C}
\def\ZPA{{\em Z. Phys.} A}

\def\EUP{{\em  Eur. Phys. J.} C}
\def\JHEP{{\em JHEP}}
\def\JPHYSG{{\em J.Phys.G}}

\newcommand{\be}{\begin{align}}
\newcommand{\ee}{\end{align}}
\newcommand{\bea}{\begin{eqnarray}}
\newcommand{\eea}{\end{eqnarray}}

\begin{document}

\title{ \vspace{1cm} On the temperature dependence of the electrical conductivity in hot quenched lattice QCD}
\author{A.\ Francis$^{1,2}$ and O. Kaczmarek$^{1}$ \\
\\
$^1$Fakult\"at f\"ur Physik, Universit\"at Bielefeld, D-33615 Bielefeld, Germany\\
$^2$Institut f\"ur Kernphysik, Johannes Gutenberg Universit\"at Mainz,\\ D-55099 Mainz, Germany
}
\maketitle
\begin{abstract} Extending our recent work \cite{Ding2010}, we report on a calculation of the vector current
correlation function for light valence quarks in the deconfined phase
of quenched QCD in the temperature range $1.16T_c\lesssim T \lesssim 2.98T_c$. 
After performing a systematic analysis of the influence of cut-off effects on light quark meson 
correlators using clover improved Wilson fermions, 
we discuss resulting constraints on the electrical conductivity in a quark gluon plasma.
\end{abstract}

%\eject
%\tableofcontents

\section{Introduction}

The spectral representation of the correlation functions of the vector
current directly relates to the invariant mass spectrum of dileptons
and photons. Additionally in the limit of small frequencies it determines a
transport coefficient, in the case of the vector correlation function
of light quarks, the electrical conductivity.\\ 
At temperatures relevant for current heavy ion experiments
non-perturbative techniques are mandatory for the determination of
these quantities.
Perturbative studies of the vector spectral functions
\cite{PTtheory,Blaizot} and also the inclusion of non-perturbative
aspects through the hard thermal loop resummation scheme
\cite{Braaten} break down, especially in the low invariant mass
region, indicated by an infrared divergent Euclidean correlator
\cite{Thoma}, leading to an infinite electrical conductivity.
However
it could be demonstrated that the
spectral function at low invariant masses in fact increases linearly
resulting in a finite electrical conductivity of the quark gluon
plasma \cite{Arnold,Gelis}. This behavior could also be established by previous work using lattice QCD \cite{AartsGupta}.\\
In \cite{Ding2010} we analyzed the vector
correlation function at $T\simeq 1.45 T_c$ and performed its
extrapolation to the continuum limit based on precise data
at various lattice sizes, corresponding to different lattice cutoffs.
We then obtained reliable results for the
determination of the spectral properties and the
extraction of the dilepton rates and transport coefficients.
Here we report on a first extension of this work to a wider temperature range 
and present results on the electrical conductivity in the temperature region $1.16T_c\lesssim T \lesssim 2.98T_c$.

\section{Thermal vector correlation and spectral function}

The desired information is encoded in the large distance region of temporal
Euclidean correlation functions of the vector current
at non-zero temperature and fixed spatial momentum,
\begin{align}
G_{\mu\nu}(\tau,\vec{p}) = \int {\rm d}^3x\ \langle J_\mu (\tau, \vec{x}) J_\nu^{\dagger} (0, \vec{0}) \rangle\ {\rm e}^{i \vec{p} \cdot \vec{x}} \; ,
\quad\textrm{where:}\quad J_\mu (\tau,\vec{x}) \equiv \bar{q}(\tau, \vec{x})\gamma_\mu q(\tau, \vec{x})\; .
\label{eq:corr_p}
\end{align}

Whereby the current-current correlation functions can be represented in terms of
an integral over spectral functions, $\rho_{\mu\nu}(\omega,\vec{p},T)$:
\begin{align}
G_{H}(\tau,\vec{p},T) =  \int_{0}^{\infty} \frac{{\rm d} \omega}{2\pi}\;
\rho_{H} (\omega,\vec{p},T)\;
\frac{\cosh(\omega (\tau - 1/2T))}{\sinh(\omega /2T)}
\quad , \quad H=00,\ ii, \ V \ ,
\label{eq:speccora}
\end{align}
here we denote by $\rho_{ii}$ the sum over the three
space-space components of the spectral function and also introduce the
vector spectral function $\rho_V \equiv \rho_{00}+\rho_{ii}$.\\

The time-time component of the vector correlation function, $G_{00}(\tau T)$, is conserved and
can be shown to be proportional to the quark number susceptibility, $G_{00}(\tau T) = - \chi_q T$.
The vector correlation functions $G_{ii}$ and $G_{V}$ therefore differ
only by a constant,
\begin{align}
G_{V} (\tau T) =  G_{ii} (\tau T) - \chi_q T \; .
\label{eq:relation}
\end{align}

At high temperature and for large energies corrections to the free field 
behavior can be calculated perturbatively; the vector spectral 
function can be deduced from the calculation of one loop corrections
to the leading order results for the thermal dilepton rate \cite{Aurenche}.
This yields,  
\begin{align}
\rho_{V}(\omega) \simeq \;
 \frac{3}{ 2 \pi} \left( 1 + \frac{\alpha_s}{\pi}  \right)
\; \omega^2  \;\tanh (\omega/4T) \quad , \quad
\omega /T \gg 1 \; .
\label{eq:spectral_ii}
\end{align}

In the free field, infinite temperature limit the spatial part of the 
spectral function contains a $\delta$-function at the origin. 
Different from the time-time component, where the $\delta$-function
is protected by current conservation, this $\delta$-function is smeared
out at finite temperature and the low energy part 
of $\rho_{ii}$ is expected to be described by a Breit-Wigner peak 
\cite{Aarts02,Teaney06,Teaney10},
\begin{align}
\rho_{ii}^{BW}(\omega) = \chi_q c_{BW}
\frac{\omega \Gamma}{\omega^2 + (\Gamma/2)^2} 
\; .
\label{eq:BWpeak}
\end{align}
The limit $\omega \rightarrow 0$ is sensitive to transport properties in the
thermal medium and in this case it leads to the 
electrical conductivity
\begin{align}
\frac{\sigma}{T} = \frac{C_{em}}{6} \lim_{\omega \rightarrow 0} 
\frac{\rho_{ii}(\omega)}{\omega T} \; .
\label{eq:conduct}
\end{align}
The Breit-Wigner form then yields $\sigma(T)/C_{em} = 2 \chi_q c_{BW}/(3 \Gamma)$. Note here that in the infinite
temperature limit the width of the Breit-Wigner peak vanishes, while at the same time 
$c_{BW}\rightarrow 1$, $\chi_q \rightarrow T^2$ and consequently the electrical 
conductivity is infinite in the non-interacting case.\\

These considerations give rise to a sensible Ansatz describing the high temperature vector spectral function:
 \begin{eqnarray}
%\rho_{00}(\omega) &=& - 2\pi \chi_q  \omega \delta (\omega)  \ ,
%\label{eq:fit00} \\
\rho_{ii} (\omega) &=&  
2\chi_q c_{BW}   \frac{\omega \Gamma/2}{ \omega^2+(\Gamma/2)^2}
+ \frac{3}{2 \pi} \left( 1 + k \right) 
\; \omega^2  \;\tanh (\omega/4T)   \ .
\label{eq:ansatz}
\end{eqnarray}
this Ansatz depends on four temperature dependent parameters; the quark number 
susceptibility $\chi_q(T)$, the strength ($c_{BW}(T)$) and width ($\Gamma (T)$) 
of the Breit-Wigner peak and the parameter $k(T)$ that parametrizes deviations 
from a free spectral function at large energies.\\
In the following we will use this Ansatz to extract the electrical conductivity
and the vector spectral function from the correlator data.

\subsection{Moments of the vector spectral function}

In addition to the vector correlation function itself we will calculate 
its curvature at the largest Euclidean time separation accessible
at non-zero temperature, {\it i.e.} at $\tau T =1/2$. The
curvature is the second 
thermal moment of the spectral functions at vanishing 
momentum,
\begin{align}
G_H^{(n)} = \frac{1}{n!} \left. \frac{{\rm d}^n G_H(\tau T)}{{\rm d} (\tau T)^n}
\right|_{\tau T =1/2}  =\frac{1}{n!} \int_{0}^{\infty} \frac{{\rm d} \omega}{2\pi}
\left( \frac{\omega}{T}\right)^n\ \frac{ \rho_H (\omega)}{{\rm sinh} (\omega /2T)}~~, \;\; H=ii,\ V\ ,
\label{eq:moments}
\end{align}
where $n$ is chosen to be even as all odd moments vanish.
These thermal moments give the Taylor expansion coefficients for the 
correlation function expanded around the mid-point of the Euclidean time 
interval,
\begin{align}
G_H(\tau T) = \sum_{n=0}^{\infty} G_H^{(2n)} \left( \frac{1}{2} - \tau T \right)^{2n} \ .
\label{eq:taylor}
\end{align}

In the infinite temperature, free field limit the integral in Eq.~\ref{eq:speccora} 
can be evaluated analytically \cite{Flo94} and one straight forwardly obtains the first three non-vanishing moments
for massless quarks
\begin{align}
G_V^{(0),free} = \frac{2}{3} G_{ii}^{(0),free} = 2 T^3 \; ,\;\; 
G_H^{(2),free} = \frac{28 \pi^2}{5} T^3
\; ,\;\; G_H^{(4),free} = \frac{124 \pi^4}{21} T^3\; .
\label{eq:free_moments}
\end{align} 
Note here that all thermal moments, $G_H^{(2n)}$ with $n\ge 0$, will be sensitive to the 
smeared  $\delta$-function contributing to $\rho_{ii}(\omega)$ although we expect this contribution 
to become more and more suppressed in higher order moments \cite{Ding2010}.\\

In the following we will analyze the ratio of $G_H(\tau T)$ and the free
field correlator $G_H^{free}(\tau T)$,
\begin{eqnarray}
\frac{G_H(\tau T)}{G_H^{free}(\tau T)} 
&=& \frac{G_H^{(0)}}{G_H^{(0),free}}
\left( 1+ \left( R^{(2,0)}_H -R^{(2,0)}_{H,free} \right) 
\left(\frac{1}{2} - \tau T \right)^{2} + ....\right) \; ,
\label{eq:series}
\end{eqnarray}
as well as the ratio of mid-point subtracted correlation functions
\begin{eqnarray}
\Delta_H(\tau T) &\equiv& 
\frac{G_H(\tau T)-G_H^{(0)}}
{G_H^{free}(\tau T)-G_H^{(0),free}}  \nonumber \\
&=& 
\frac{G^{(2)}_H}{G^{(2),free}_{H}}
\left( 1+ \left( R^{(4,2)}_H -R^{(4,2)}_{H,free} \right)
\left(\frac{1}{2} - \tau T \right)^{2} + ....\right) \; .
\label{eq:mid-point}
\end{eqnarray}
Here we used the notation $R_H^{(n,m)} \equiv G_H^{(n)}/G_H^{(m)}$.
Note that the curvature of these ratios at the mid-point determines the 
deviation of ratios of thermal moments from the
corresponding free field values. While the ratios of 
correlation functions differ in the $H=ii$ and $H=V$ channels due to 
the additional constant contributing to $G_{V}(\tau T)$, this constant 
drops out in the subtracted correlation function, {\it i.e.}, 
$\Delta_V(\tau T)\equiv \Delta_{ii}(\tau T)$ and $G_V^{(n)}= G_{ii}^{(n)}$
for $n>0$. 

\begin{table}[b!]
\begin{center}
\begin{tabular}{|c|c|c|c|c|c|c|c|c|c|}
\hline
$N_\tau$ & $N_\sigma$ & $\#$ conf &$\beta$ & $a[\textrm{fm}]$ &$T/T_c$ & $c_{SW}$ & $\kappa$ & $m_{\overline{MS}}/T_{[\mu=2\textrm{{\tiny GeV}}]}$ &$Z_V$ \\\hline
40& 128 & 451 & 7.457 & 0.015 &1.16 & 1.3389 & 0.13390 & 0.0989(4)&0.851\\
32& 128 & 255 & 7.457 & 0.015 &1.49 & 1.3389 & 0.13390 & 0.0989(4)&0.851\\
16& 128 & 340 & 7.457 & 0.015 &2.98 & 1.3389 & 0.13390 & 0.0989(4)&0.851  \\\hline
48& 128 & 191 &7.793 & 0.010 &1.45 & 1.3104 & 0.13340 & 0.1117(2)&0.861 \\\hline
\end{tabular}
\end{center}
\caption{Parameters for the calculation of vector correlation functions on
lattices of size $N_\sigma^3\times N_\tau$.}
\label{all-par}
\end{table}

\section{Temperature dependence of the vector spf on the lattice}
%\section{Electrical conductivity and thermal dilepton rate}

%\subsection{Computational Setup}

The numerical results presented here are an extension to our earlier study and comprise calculations that
have been performed at several temperatures $1.16T_c\lesssim T\lesssim 2.98T_c$. Consequently all parameters
were chosen and adjusted following the procedures outlined in \cite{Ding2010}.
As such they are obtained from quenched QCD gauge field configurations generated with the standard SU(3) 
single plaquette Wilson gauge action \cite{wilson} using the clover improved Wilson action with non-perturbatively chosen clover 
coefficient $c_{SW}$ \cite{luescher1} in the fermion sector.
The temperature range is then varied by keeping the gauge coupling fixed
for different temporal lattice sizes $N_\tau = 16$, 32, 40. Note here throughout 
we also show results at $T\simeq1.45T_c$ for $N_\tau=48$ for comparison. 
The simulation parameters are given in Tab.~\ref{all-par}.\\

%\subsection{Numerical Results}

\begin{figure}[t!]
\includegraphics[width=.45\textwidth]{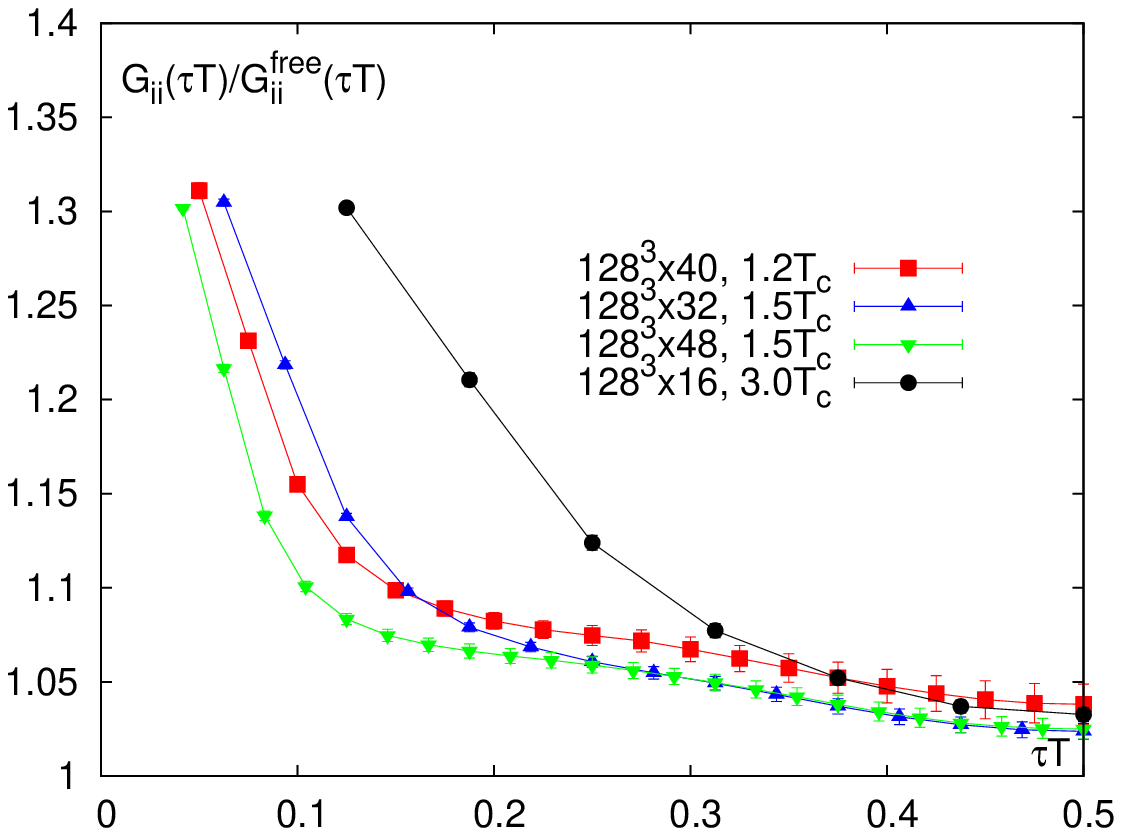}
\includegraphics[width=.45\textwidth]{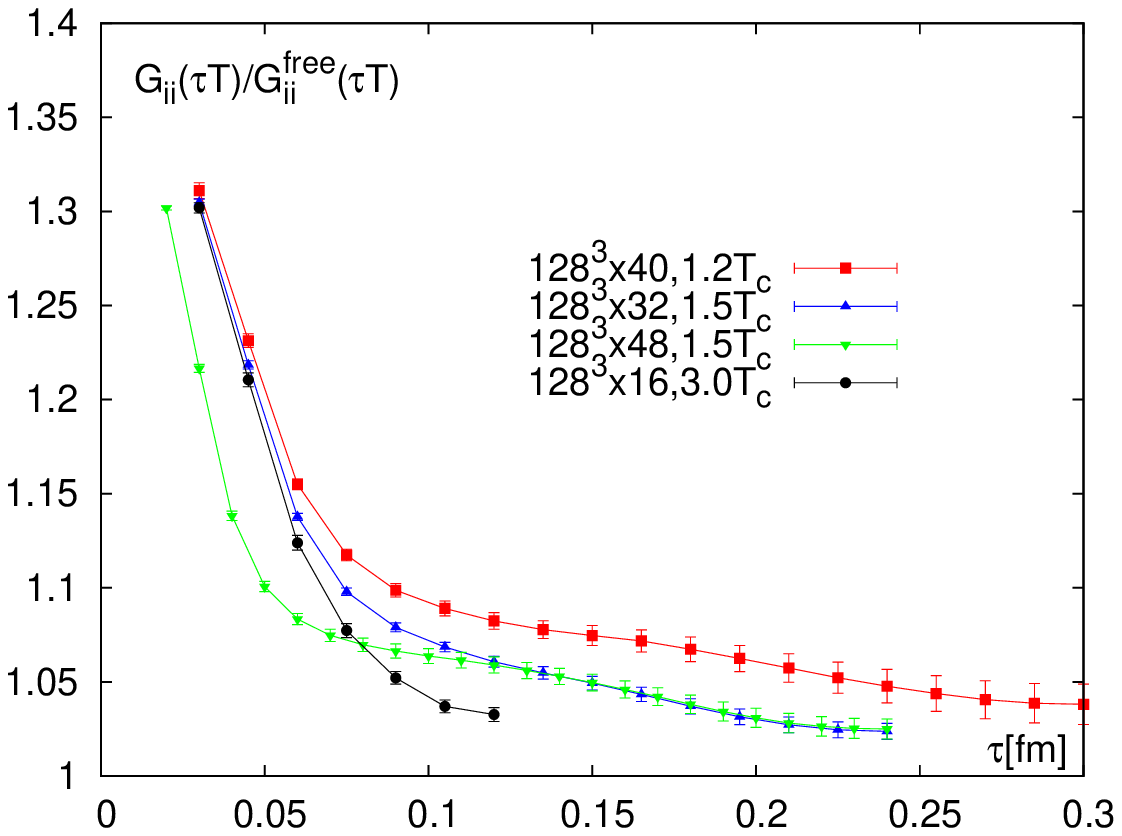}
\caption{The vector correlation function normalized by its free continuum counterpart at varying temperature
in $\tau T$ (left) and [fm] (right).
}
\label{fig:corr-tdep}
\end{figure}

In Fig.\,\ref{fig:corr-tdep} 
we show the ratio $G_{ii}(\tau T)/G_{ii}^{free}$ for the $\beta=7.457$ and the $\beta=7.793$ (green points) lattices
in Euclidean temperature units (left) and in physical distance (right).
On the right of Fig.\,\ref{fig:corr-tdep} we see that the $\beta=7.457$ lattices lie on top of each other 
for $\tau\lesssim0.06$fm, in view of the results at $\beta=7.793$ this can be understood as due to cut-off effects. 
As such the cut-off effects dominate the low distance region
of the correlation function, in the presented calculations however the cut-off is fixed and therefore
we expect them to be the same throughout.
At $\tau\gtrsim0.06$fm the deviations should subsequently be due to
temperature effects. Notice however that the first 6 to 8 points in the correlator are
dominated by the cut-off and consequently the $N_\tau=16$ results are more affected by the
cut-off effects throughout the Euclidean time interval.
Nevertheless the $N_\tau=16$ results in Fig.\,\ref{fig:corr-tdep} (left) level out in the regime $\tau T\gtrsim0.4$.
Note the correlator ratio from the $T\simeq1.16T_c$ lattice exhibits a visibly larger value
for all Euclidean times, even though its trend is similar to that at $T\simeq1.49T_c$. 
%This could originate from a much broader transport peak that subsequently leaves a larger imprint on the correlator 
%at large distances or it could be a combination of both the transport peak and a possible $\rho$-resonance 
%contribution.\\

\subsection{Temperature dependence of the thermal moments}

\begin{figure}
\begin{minipage}[l]{0.45\textwidth}
 \includegraphics[width=\textwidth]{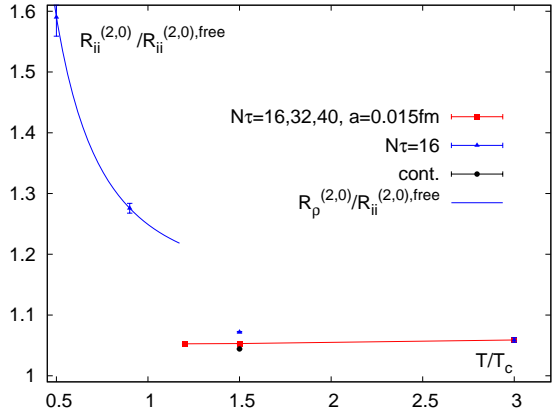}
\end{minipage}
\hspace{0.1cm}
\begin{minipage}[l]{0.25\textwidth}
  \begin{tabular}{c c c c} \hline\hline
$N_\tau$ & $T/T_c$ & $R_{ii}^{(2,0)}/R_{ii}^{(2,0),free}$ \bigstrut\\ \hline\bigstrut[t]
16 & 2.98 & 1.0589(39)  \\
32 & 1.49 & 1.0530(4) \\
40 & 1.16 & 1.0531(5)\\\hline\bigstrut[t]
16 & 1.50 & 1.0718(12)\\
16 & 0.93  & 1.2757(80)\\
16 & 0.55  & 1.590(30)\\\hline\hline
   \end{tabular}
\end{minipage}
\caption
{Left: The ratio $R_{ii}^{(2,0)}/R_{ii}^{(2,0),free}$ over $T/T_c$ above and below $T_c$. Shown are
the results for $\beta=7.457$ (red) and the continuum extrapolation at $T\simeq1.45T_c$ (black). Additionally
a number of results from $N_\tau=16$ lattices at varying temperatures (blue) are given, 
including also results that have been obtained using data from \cite{Wissel}. Right: Table of results
for $R_{ii}^{(2,0)}/R_{ii}^{(2,0),free}$.} 
\label{fig:thmom-tdep2}
\end{figure}

Calculating the temperature dependence of the thermal moments we compute
the quantity $\Delta_H(\tau T)$ and then fit to a quartic Ansatz. Using a jackknife procedure
we then immediately calculate the ratio $R_{ii}^{(2,0)}/R_{ii}^{(2,0),free}$ and the corresponding 
results over $T/T_c$ are shown in Fig.\,\ref{fig:thmom-tdep2}. 
Note here we also give the $N_\tau=16$ and continuum extrapolated
results at $T\simeq1.45T_c$ taken from \cite{Ding2010} for comparison.
Additionally here we also show points below $T_c$ from calculations done by the Bielefeld lattice group that have
already been examined in \cite{Wissel}.\\ 
Above $T_c$, i.e. in temperature region evaluated here $1.16T_c\lesssim T \lesssim 2.98T_c$,
the data points are seen to be almost constant in $T/T_c$. Turning to the table
on the right of Fig.\,\ref{fig:thmom-tdep2} we identify a slight rise of the ratio within $0.5\%$
going from $\simeq1.49T_c$ and $N_\tau=32$ to $\simeq2.98T_c$ and $N_\tau=16$.\\
Below $T_c$ on the other hand we immediately identify a clear temperature dependence of the ratio 
$R_{ii}^{(2,0)}/R_{ii}^{(2,0),free}$. As such there is a $30\%$ difference between the results at $T\simeq0.55T_c$
and $T\simeq0.93T_c$. At the same time the result at $T\simeq0.93T_c$ is roughly $20\%$ larger than that encountered
above $T_c$.\\

Together the drop below and the constant behavior above $T_c$ are very interesting from the point of
view of a possible $\rho$-resonance in the spectrum. As such we estimate the impact of a $\rho$-resonance
on the ratio of thermal moments $R_{ii}^{(2,0)}/R_{ii}^{(2,0),free}$ by assuming 
the $\rho$-resonance to contribute a single $\cosh$ in the correlator:
\begin{align}
 G_{\rho}(\tau)=A\cdot\cosh\Big[\frac{m_\rho}{T}\cdot \tau T\Big]\quad,
\end{align}
taking the second derivative in $\tau T$ of $G_\rho(\tau)$ then gives the second thermal moment, thus
we find:
\begin{align}
 R_\rho^{(2,0)}=\Big( \frac{m_\rho}{T}\Big)^2=\Big( \frac{m_\rho}{T_c}\Big)^2\cdot\Big( \frac{T_c}{T}\Big)^2\quad.
\end{align}
As a consequence we expect the ratio $R_\rho^{(2,0)}$ to go like $\sim1/T^2$. Adding also a continuous contribution
to this model ratio we arrive at the following estimate for the temperature dependence below $T_c$:
\begin{align}
 R_{T<T_c}^{(2,0)}= c_{cont} + \Big( \frac{c_{\rho}m_\rho}{T_c}\Big)^2\cdot\Big( \frac{T_c}{T}\Big)^2\quad.
\end{align}
Even though we have only two points below $T_c$ we can nevertheless fit this estimate to the data and the
result is shown as blue line in Fig.\,\ref{fig:thmom-tdep2}.
%The mass contribution of the $\rho$-resonance
%calculated in this way is $c_\rho m_\rho(T<T_c)\simeq1.67$GeV.\\
The qualitatively very different behavior above and below $T_c$,
subsequently leads us to conclude that the drop of the ratio $R_{ii}^{(2,0)}/R_{ii}^{(2,0),free}$ below $T_c$
is in fact mostly due to a particle contribution.
At the same time the constant behavior of the ratio above $T_c$
suggests that the $\rho$-resonance does not or only very weakly contribute
in the examined temperature range.

\subsection{Consequences for the vector spectral function}

\begin{table}[t]
\begin{center}
%\vspace{0.3cm}
\begin{tabular}{l l c c c c c }\hline\hline
$T/T_c$ & $\tau_{min} T$ & $2c_{BW}\widetilde{\chi}_q/\widetilde{\Gamma}$ & $\widetilde{\Gamma}$ & $k/\widetilde{\chi}_q$ & $\chi^2/dof$ &
$1/C_{em}\cdot\sigma/T$ \bigstrut\\ \hline\bigstrut[t]
2.98 & 0.396  & 0.93(19) & 3.07(70) & 0.160(19) & 0.54 & 0.31(7)\\
     & 0.354  & 0.91(11) & 2.93(78) & 0.166(11) & 0.72 & 0.30(4)\\
     & 0.3125 & 0.90(8)  & 3.08(36) & 0.189(6)  & 1.37 & 0.30(3)\\
     & 0.25   & 0.71(2)  & 3.75(14) & 0.122(4)  & 2.07 & 0.24(1)\\\hline
1.49 & 0.396  & 0.98(23) & 3.14(80) & 0.165(23) & 0.01 & 0.33(8)\\
     & 0.354  & 0.97(11) & 3.16(40) & 0.166(11) & 0.07 & 0.32(4)\\
     & 0.3125 & 0.92(4)  & 3.47(16) & 0.164(7)  & 0.26 & 0.31(1)\\
     & 0.25   & 0.90(2)  & 3.53(10) & 0.170(4)  & 1.16 & 0.30(1)\\\hline
1.16 & 0.396  & 1.00(34) & 3.25(12) & 0.235(38) & 0.01 & 0.33(11)\\
     & 0.354  & 0.99(17) & 3.30(62) & 0.236(18) & 0.01 & 0.33(6)\\
     & 0.3125 & 0.88(6)  & 3.89(20) & 0.233(12) & 0.05 & 0.29(7)\\
     & 0.25   & 0.87(4)  & 3.89(18) & 0.242(6)  & 0.14 & 0.29(1)\bigstrut[b]\\\hline\hline
\end{tabular}
\end{center}
\caption
{The fit parameters of the Breit-Wigner+continuum Ansatz for $T\simeq2.98T_c,\, 1.49T_c$ and $T\simeq1.16T_c$ on lattices with temporal extent
$N_\tau=16,\,32$ and $40$. Note the fit window is varied between $\tau_{min} T=0.25$ and $\tau_{min} T=0.396$. 
%When the corresponding point was not available on the lattice at hand a spline interpolation combined with a jackknife analysis
%was used to provide it
.
}
\label{tab:fit-tdep}
\end{table}

To discuss the consequences of our findings on the vector spectral function
we invoke the Breit-Wigner+continuum Ansatz defined in Eq.\,\ref{eq:ansatz} and fit 
to the $T\simeq1.16T_c,\, 1.49T_c$ and $T\simeq2.98T_c$ data respecting also the thermal moments. 
Additionally we analyze the dependence of the fit parameters on $\tau_{min} T$, 
as, with the data being subject to potentially large lattice 
effects, we expect large errors originating from the fit-window we choose. 
The resulting parameters are summarized in Tab.\,\ref{tab:fit-tdep}.\\
Here it can be seen that the results at $\tau_{min} T=0.396$ and $\tau_{min} T=0.354$ 
have the largest errors,
nevertheless these results should be closest to those in the continuum
as only the furthest distance points are comparatively free of lattice effects. Without
a continuum extrapolation however it is at this point not possible to cleanly discern the residual deviation
due to the lattice cut-off.
Decreasing $\tau_{min} T$ on the other hand leads to visible
systematic trends in the parameters, as such the correction term $k/\widetilde{\chi}_q=kT^2/\chi_q$ 
and the width $\widetilde{\Gamma}=\Gamma/T$ both increase with decreasing $\tau_{min} T$, while the electrical conductivity decreases.
Additionally the $\chi^2/dof$ increases with decreasing $\tau_{min} T$ implying a lower quality fit
with lower $\tau_{min} T$. 
In light of the argument above these results show that indeed the lattice effects influence
the fit more strongly with smaller $\tau_{min} T$, as they cannot be taken care of by the employed Ansatz.
Consequently we attribute the largest part of the systematic trends observed above to the lattice effects.\\
Focusing on the parameters themselves note that the correction factor $k(T)$ is the most accurately determined. 
Clearly it shows an increasing trend with decreasing temperature differing by a factor $\sim1.5$ between
$T\simeq2.98T_c$ and $T\simeq1.16T_c$. However as $k(T)$ is coupled to the strong coupling via $k(T)\simeq \alpha_s/\pi$
\cite{Aurenche}, this behavior is expected.
The same can be seen for the width $\widetilde{\Gamma}$ and 
the parameter $2c_{BW}\widetilde{\chi}_q/\widetilde{\Gamma}=2c_{BW}\chi_q/(\Gamma T)$, they also slightly increase with decreasing temperature.
However here, they remain within errors of each other and the increase between $T\simeq2.98T_c$ and $T\simeq1.16T_c$
is small.
Naturally this also leads to only small deviations of the electrical conductivity. 
Note however that the results at $N_\tau=16$ differ by $\simeq5\%$ from those of $N_\tau=32$, while
this deviation is only $\simeq2\%$ for $N_\tau=40$. With the $N_\tau=16$ lattice being most affected
by the systematics of the fit, we assume the largest part of this deviation to find its
origins there. 
Even so we find the electrical conductivity to be to a good degree constant 
and within $\simeq7\%$ across the temperature region evaluated here, whereby
the values at the individual temperatures have fit errors around $\simeq(7-11)\%$.\\
In the temperature range $1.16T_c\lesssim T \lesssim 2.98T_c$ Our results therefore imply that the  
electrical conductivity depends linearly on temperature as expected from perturbation theory \cite{Arnold} and at $\beta=7.457$
we find the value:
\begin{align}
 \sigma= (0.33\pm 0.04 \pm0.11)\cdot C_{em}\cdot T\quad,
\end{align}
whereby the first error quoted corresponds to the temperature effects and the second to
the fit errors.

\section{Conclusions}

Extending our earlier study we presented results on the 
vector correlation function in the high temperature phase of quenched QCD in
the temperature range $1.16T_c\lesssim T \lesssim 2.98T_c$.
First results show a similar behavior to those extensively analyzed at $T\simeq 1.45T_c$ and
even at $T\simeq1.16T_c$ we find no sizable contribution attributable to the $\rho$-resonance.\\
Even though these calculations could not yet be extended to include also a continuum extrapolation
as in \cite{Ding2010}, we were nevertheless able to extract the relevant spectral properties to calculate the
electrical conductivity via an Ansatz and found this phenomenologically relevant transport coefficient to depend 
linearly on temperature with a value of $\sigma= (0.33\pm 0.04 \pm0.11)\cdot C_{em}\cdot T$.\\
Given the exact relation between the electrical conductivity and the diffusion constant, $\sigma= \chi_q\,D$,
it would be interesting to compare these results with those obtained in the case of charm diffusion \cite{HQD} or from the
momentum diffusion constant $\kappa$ calculable via HQET \cite{HQET,HQET2} and we will do so in an upcoming publication.

\end{document}